\documentclass{revtex4}
\usepackage{bm}
\usepackage{hyperref}
\usepackage{color}
\usepackage{latexsym}
\usepackage{amscd,amsmath,amssymb}

\newcommand{\p}[1]{(\ref{#1})}

\newcommand{\bG}{{\overline G}{}}

\newcommand{\bp}{{\bar p}}

\newcommand{\bu}{{\bar u}}

\newcommand{\bx}{{\bar x}}

\newcommand{\bv}{{\bar v}}

\newcommand{\sfrac}[2]{{\textstyle\frac{#1}{#2}}}

\newcommand{\be}{\begin{equation}}
\newcommand{\ee}{\end{equation}}
\newcommand{\bea}{\begin{eqnarray}}
\newcommand{\eea}{\end{eqnarray}}
\newcommand{\ba}{\begin{array}} \newcommand{\ea}{\end{array}}

\def\im{{\rm i}}
\newcommand{\nn}{\nonumber}

\begin{document}
\begin{flushright}
\end{flushright}
\title{SU(1,2) invariance in two-dimensional oscillator}
\author{Sergey Krivonos}
\email{krivonos@theor.jinr.ru}
\affiliation{Bogoliubov Laboratory of Theoretical Physics, Joint Institute for Nuclear Research, 141980 Dubna, Russia}
\author{Armen Nersessian}
\email{arnerses@ysu.am}
\affiliation{Yerevan State University, 1 Alex Manoogian St., Yerevan, 0025, Armenia}
\affiliation{Tomsk Polytechnic University, Lenin Ave. 30, 634050 Tomsk, Russia}

\begin{abstract}
Performing the Hamiltonian analysis we explicitly established the canonical equivalence of the deformed
oscillator, constructed in {\tt arXiv:1607.03756}, with the ordinary one. As an immediate consequence, we proved that the $SU(1,2)$
symmetry is the dynamical symmetry of the ordinary two-dimensional oscillator. The characteristic feature of this
$SU(1,2)$ symmetry is a non-polynomial structure of its generators written it terms of the oscillator variables.
\end{abstract}

\maketitle
\setcounter{page}{1}
\section{Introduction}
It is a well-known fact that the invariance with respect to the $\ell >1/2$-conformal Galilei algebra \cite{confNH1,confNH2}
demands the appearance of high-derivative terms in the Lagrangians of the corresponding mechanical systems \cite{confNH3,confPU,GM1,Ivanov,MT,PU11}. The important fact is that standard methods of nonlinear realizations \cite{coset1,coset2} work quite nicely for these algebras  being equipped with the Inverse Higgs phenomenon constraints \cite{ih} result in the corresponding Pais-Uhlenbeck oscillators \cite{KLS}. The exceptional case with $\ell=1/2$ corresponds to the Shr\"{o}dinger algebra, and the mechanical system possessing this symmetry is just a standard $d$-dimensional oscillator. It was demonstrated in the recent paper \cite{KLS} that the $su(1,2)$ algebra admits a reduction to the two-dimensional Shr\"{o}dinger algebra and, therefore, the system possessing the $SU(1,2)$ symmetry reduces to the ordinary two-dimensional oscillator. Such deformed oscillator has been constructed in \cite{KLS} within the Lagrangian formalism.

As for a possible relation of the deformed oscillator with the ordinary one, one should note that it seems to be impossible
to relate these systems within the Lagrangian approach. Contrary, within the Hamiltonian approach the freedom to relate
these systems is much wider, because the admitted change of variables includes arbitrary (but invertible) functions defined on the phase-space.
That is why we provide a Hamiltonian description of the $su(1,2)$ oscillator in the present paper. It turns out that the
standard procedure to pass to the Hamiltonian formalism is not much suitable for the present case resulting in a rather complicated
Hamiltonian. The basic explanation of this is that the canonically defined momenta have rather complicated transformation properties
with respect to the  $SU(1,2)$ group. On the other hand, within the nonlinear realization approach applied to this system \cite{KLS},
we have proper variables $v, \bv$ in the coset space which can be used as  proper momenta with transparent transformation properties.
Interestingly enough, one of the Cartan forms, used as the Lagrangian in \cite{KLS}, is capable of providing  the symplectic structure
as well as the Hamiltonian in terms of the initial variables $u,\bu,v,\bv$. The complicated structure of the Poisson brackets in this basis is
compensated by the simple form of the Hamiltonian and  the generators of $su(1,2)$ algebra.

Having at hand all ingredients in the initial variables, we succeeded in finding the canonical variables in which the
Hamiltonian of deformed $su(1,2)$-invariant oscillator coincides with the Hamiltonian of the ordinary two-dimensional oscillator.
Thus, we proved that these two systems are canonically equivalent. However, in these canonical variables the generators of  $su(1,2)$ algebra have a non-polynomial structure, so it is problematic to state about their quantum equivalence.

\setcounter{equation}{0}
\section{Deformed oscillator in the Lagrangian approach}
In  \cite{KLS}, the Lagrangian of the deformed oscillator
\be\label{Lag1}
\mathcal{L} = \frac{ \dot{u} \dot{\bu} - \omega^2 u \bu}{1+ \frac{\im \gamma}{4} \left( u \dot{\bu}- \dot{u} \bu\right)+
\frac{\gamma^2 \omega^2}{4} u^2\, \bu{}^2}
\ee
was constructed within nonlinear realization of the $SU(1,2)$ group. The structure relations of the corresponding algebra $su(1,2)$ were
chosen as
\be
\label{w32}
\begin{gathered}
  \im \left[ L_n, L_m\right] = (n-m) L_{n+m}, \;  \im \left[L_n, G_r \right] =\left( \frac{n}{2}-r\right) G_{n+r}, \;\im
\left[L_n, \bG_r \right] =\left( \frac{n}{2}-r\right) \bG_{n+r},  \\
  \left[U, G_r\right] =  G_r, \quad \left[U, \bG_r\right] =  - \bG_r,\\
 \im \left[ G_r, \bG_s \right] = \gamma \left( \sfrac{3}{2} (r-s) U - \im  L_{r+s}\right), \quad n,m=-1,0,1,\; r,s = -1/2, 1/2.
\end{gathered}
\ee
In this form, in the limit $\gamma=0$ these relations  coincide with the relations of the  $\ell=\frac{1}{2}$ conformal Galilei algebra in three
dimensions \cite{confNH1,confNH2}, and so they can be viewed as the deformation of the conformal Galilei algebra with the parameter of deformation  $\gamma$.
The exact value of $\gamma$ is inessential: if nonzero, it can be put to unity by a re-scaling of the generators $G_r$ and $\bG_r$.

The group $SU(1,2)$ itself was realized by the left multiplication of the coset $g=SU(1,2)/H$ with the stability subgroup $H\propto(U, L_0, L_1)$
parameterized as
\be\label{coset33}
g=e^{\im t \left( L_{-1}+\omega^2 L_1\right)}\, e^{\im \left(u G_{-1/2}+\bu \bG_{-1/2}\right) }\,  e^{\im \left(v G_{1/2}+\bv \bG_{1/2}\right)}.
\ee
Using the Cartan forms defined in a standard way as
\bea
&& g^{-1}\, d\, g =  \im \, \sum_{n=-1}^1 \Omega_{n}  L_{n}  + \im \, \sum_{\alpha=-1/2}^{1/2}
\left( \omega_\alpha G_\alpha +{\bar\omega}_\alpha \bG_\alpha\right) +
\im \, \Omega_U U,   \label{CF}
\eea
one may eliminate the unessential Goldstone fields $v, \bv$ via the fields $u, \bu$ by imposing the constraints \footnote{This is the particular case of the Inverse Higgs phenomenon conditions \cite{ih}.}
\be\label{ih34}
\omega_{-1/2}={\bar\omega}_{-1/2}=0\qquad \Rightarrow \qquad v = \frac{\dot{u}+\im \frac{\gamma \, \omega^2}{2} u^2\, \bu}{1+\im \frac{\gamma}{2}
\left( u \dot{\bu} - \bu \dot{u}\right) +\frac{\gamma^2 \, \omega^2}{4} u^2\, \bu^2},\;
\bv = \frac{\dot{\bu}-\im \frac{\gamma \, \omega^2}{2} u\, \bu{}^2}{1+\im \frac{\gamma}{2} \left( u \dot{\bu} - \bu \dot{u}\right) +
\frac{\gamma^2 \, \omega^2}{4} u^2\, \bu^2}.
\ee
The action for the deformed oscillator is provided by the Cartan form $\Omega_U$ \p{CF}, which explicitly reads \cite{KLS}
\bea\label{CFU}
\Omega_U & = & \sfrac{3}{2} \gamma \left[ v\, \bv\, \Big( \left( 1+ \sfrac14 \gamma^2 \, \omega^2 u^2\, \bu{}^2\right) dt+ \sfrac{\im}{2}\gamma\left( u\, d \bu - \bu\, du\right)\Big) -  \right. \nn \\
&&  \left. v \left( d \bu  -\sfrac{\im}{2}\gamma\,\omega^2 u \, \bu^2\, dt\right) -
\bv\, \left(d u + \sfrac{\im}{2}\gamma\,\omega^2 u^2 \, \bu\, dt\right) +\omega^2 u\, \bu\,dt \right].
\eea
Finally, upon the substitution of \p{ih34} into \p{CFU} one may get
\be\label{act1}
S=  -\sfrac{2}{3 \gamma} \int \Omega_U = \int  \mathcal{L}\,dt
\ee
where the Lagrangian is given by \eqref{Lag1}.
One has to stress again, that the action \p{act1} is invariant with respect to $SU(1,2)$ symmetry.

Finally, note that the action
\be\label{aac}
S=  -\sfrac{2}{3 \gamma} \int \Omega_U
\ee
with the form $\Omega_U$ given by expression \p{CFU} is sufficient to describe the deformed oscillator without any references to the
inverse Higgs constraints \p{ih34}. Indeed, varying the action \p{aac} over the variables $v, \bv$ we immediately reproduce the
constraints \p{ih34}. Thus, the action \p{aac} contains all needed information to describe the deformed oscillator.
\setcounter{equation}{0}
\section{Hamiltonian formulation}
To provide the Hamiltonian description of the deformed oscillator with the Lagrangian \eqref{Lag1}, one may perform  Legendre transformation  and get the system with canonical Poisson brackets with the momenta $\pi, \bar\pi$ canonically conjugated with $u,\bu$ variables. However, the Hamiltonian system written in these terms not very convenient for further analyzes.

Interestingly enough, the non-linear realization approach allows to give Hamiltonian formulation of the system in suitable phase space coordinates, without referring to Legendre transformation.   The key observation is that
the form $\Omega_U$ \p{CFU} provide us with first-order Lagrangian which is variationally equivalent to \eqref{Lag1}:
$$
\widetilde{\mathcal{L}}\, dt = -\sfrac{2}{3 \gamma}  \Omega_U = \bm{\alpha} -  \mathcal{H}\,dt ,
$$
where
\be
\bm{\alpha}= v d \bu+ \bv du + \imath \frac{\gamma}{2} v \bv \left( \bu du - u\, d\bu\right), \label{A}\ee
is  the symplectic one-form and
\be
\mathcal{H} = v\, \bv +\omega^2 \,u \,\bu \left(1+ \frac{\im}{2} \gamma \, \bu\,v\right)\left(1- \frac{\im}{2} \gamma \, u\,\bv\right). \label{H}
\ee
is the Hamiltonian.
The external differential of the symplectic one-form yields the symplectic structure
\be\label{ss}
\Omega = d \bm{\alpha} = \left( 1- \frac{\im \gamma}{2}  u\, \bv \right) dv \wedge d\bu+\left( 1+ \frac{\im \gamma}{2}  \bu\, v \right) d\bv \wedge du +\frac{\im \gamma}{2} \left( \bu\, \bv\, dv \wedge du - u\, v\, d\bv \wedge d\bu\right) +\im \gamma\, v\, \bv\, d\bu \wedge du.
\ee
The respective Poisson brackets are defined by the following non-zero relations variables $u, \bu, v, \bv$
\be\label{PB}
\{v,\bu \}=\frac{1+\imath\frac{\gamma}{2}v\bu}{1-\imath\frac{\gamma}{2}(u\bv-\bu{v})}, \qquad \{v,\bv \}=-\imath\gamma \frac{v\bv}{1-\imath\frac{\gamma}{2}(u\bv-\bu{v})} \qquad
\{v,u\}=\frac{\imath\frac{\gamma}{2}\,v u}{1-\imath\frac{\gamma}{2}(u\bv-\bu{v})},
\ee
and their complex conjugated ones.
%
Let us notice that from the  \eqref{A} one can immediately get
the expression the canonical momenta $\pi, \bar\pi$  which arise  within Legendre transformation of the Lagrangian \eqref{Lag1}
\be
\pi=v-\imath\frac{\gamma}{2}v\bv u,\quad \bar\pi=\bv+\imath\frac{\gamma}{2}v\bv \bu\; :\qquad \{\pi, \bu\}=\{\bar\pi,u\}=1.
\label{can}\ee

To complete this Section, let us  write down, in terms of $u, {\bar u}, v, \bv$,  the Hamiltonian realization of the  $su(1,2)$  generators:
\bea\label{realis0}
&&L_{-1}= v\, \bv,\quad  L_0  =-\frac{1}{2} \left(u\,\bar{v}+\bar{u}\, v\right),\quad
L_1 =u \,\bu \left(1+ \frac{\im}{2} \gamma \, \bu\,v\right)\left(1- \frac{\im}{2} \gamma \, u\,\bv\right), \\
&&U= \im \left(\bar{u}\,v - u\, \bar{v}\right)+\gamma\, u\,\bu\,v\,\bv,
\quad G_{-1/2}=-\bv \left( 1 + \im \gamma\, \bu\,v\right), \quad \bG_{-1/2}=-v \left( 1 - \im \gamma\, u\,\bv\right),\label{simfree}\\
&& G_{1/2}=\bu \left( 1 + \im \gamma\,\bu\,v\right)\left(1 -\frac{\im}{2} \gamma \,u \bv\right),\quad
\bG_{1/2}=u \left( 1 - \im \gamma\,u\,\bv\right)\left(1 +\frac{\im}{2} \gamma \,\bu v\right) .
\eea
These generators form the $su(1,2)$ algebra with respect to the  Poisson brackets  \p{PB}
\be\label{w32pb}
\begin{gathered}
   \left\{ L_n, L_m\right\} = (n-m) L_{n+m}, \quad   \left\{L_n, G_r \right\} =\left( \frac{n}{2}-r\right) G_{n+r}, \quad
\left\{L_n, \bG_r \right\} =\left( \frac{n}{2}-r\right) \bG_{n+r},  \\
  \left\{U, G_r\right\} = \im\, G_r, \quad \left\{U, \bG_r\right\} =  -\im \,\bG_r,\quad
  \left\{ G_r, \bG_s \right\} =  - \im \gamma L_{r+s}+ \sfrac{3}{2} \gamma\, (r-s) \left(U+\frac{2}{3\gamma}\right),
\end{gathered}\ee
where $n,m=-1,0,1,\; r,s = -1/2, 1/2$.

It should be noted the appearance of the constant central charge in the Poisson brackets $\left\{ G_r, \bG_s \right\}$. If $\gamma\neq 0$,
it can be absorbed by redefinition of the generator $U \rightarrow {\widetilde U}= U+\frac{2}{3\gamma}$. But if $\gamma=0$, this central
charge survives and we have at hand the central charge extension of the $\ell=1/2$ conformal Galilei algebra.

Time-dependent extensions of the generators,  defining the isometries of the Lagrangian of the deformed oscillator, are given by the expressions
\be
\begin{aligned}
&L^{t}_{-1}=\cos^2( \omega t)\left(L_{-1}+\omega^2 L_1\right) - \omega\sin( 2\omega t)\; L_0- \omega^2 \cos(2\omega t)\;L_1,
\\
& L^{t}_0=\cos( 2\omega t)\; L_0+\frac{\sin( 2\omega t)}{2\omega} \left(L_{-1}-\omega^2 L_1\right),
\\
&L^{t}_{1}=\frac{\sin^2( \omega t)}{\omega^2}\left( L_{-1}+\omega^2 L_1\right) +\cos( 2\omega t) \; L_1 +\frac{\sin( 2\omega t)}{\omega}\;L_0,\quad U^t=U,
\\
&G^{t}_{-1/2}=\cos( \omega t) \;G_{-1/2} -\omega\, \sin(\omega t)\; G_{1/2}, \quad
\bG^{t}_{-1/2}=\cos( \omega t) \;\bG_{-1/2} -\omega\, \sin(\omega t)\; \bG_{1/2},
 \\
&G^{t}_{1/2}= \cos( \omega t)\; G_{1/2} + \frac{(\sin \omega t)}{\omega} \;G_{-1/2}, \quad
\bG^{t}_{1/2}= \cos( \omega t)\; \bG_{1/2} + \frac{(\sin \omega t)}{\omega} \;\bG_{-1/2}.
\end{aligned}
\ee

\setcounter{equation}{0}
\section{Canonical variables}
Within the Hamiltonian description of the given system, we have a much more possibilities to redefine the variables than in
the Lagrangian approach. In this Section we will demonstrate that the deformed oscillator with the Hamiltonian \p{H} and
the symplectic structure \p{ss} is canonically  equivalent to the ordinary oscillator. To simplify the presentation, we  start
with the deformed free particle (i.e. with $\omega=0$) and then will consider the deformed oscillator in a full generality.
\subsection{Free particle}
In the free particle case, i.e. when $\omega=0$, the  Hamiltonian is given by the generator $L_{-1}$
\be
\label{H1}\mathcal{H}_0 = v \bv.
\ee
It has three constants of motion given by the  generators $G_{-1/2}, \bG_{-1/2}$ and $U$ \eqref{simfree}
\be
\{G_{-1/2}, \mathcal{H}_0\} =  \{\bG_{-1/2}, \mathcal{H}_0\} = \{U, \mathcal{H}_0\}=0
\ee
It immediately follows from relations \p{w32} that
\be\label{comm1}
\left\{ G_{-1/2}, \bG_{-1/2}\right\} =- \im \gamma L_{-1} = - \im \gamma \mathcal{H}_0,\quad \left\{ U, G_{-1/2}\right\}= \im\, G_{-1/2},\;
\left\{ U, \bG_{-1/2}\right\}= - \im\, \bG_{-1/2}.
\ee
The Hamiltonian can be written down in terms of these constants of motion
\be\label{H2}
\mathcal{H}_0 = \frac{G_{-1/2} \, \bG_{-1/2}}{1+\gamma U} 
\ee
It is slightly unexpected that the evident definitions of the new variables $p, \bp$
\be\label{pp}
p= -\frac{G_{-1/2}}{\sqrt{1+\gamma\, U}},\quad \bp= -\frac{\bG_{-1/2}}{\sqrt{1+\gamma\, U}},\qquad \mathcal{H}_0 = p\, \bp
\ee
provide us with proper momenta because $\left\{ p, \bp \right\} =0$.
To get complete correspondence with free particle, we have to find the  coordinates $x$, $\bar{x} $  canonically conjugated with the momenta $p, \bp$.
Explicitly, they read
\bea
&& x=\bu\;\frac{2+ \gamma\, U+ \im\,\gamma\, \bu\,v}{2 \sqrt{1+\gamma\,U}} ,\quad
{\bar x}=u\;\frac{2+ \gamma\, U- \im\,\gamma\, u\,\bv}{2\sqrt{1+\gamma\,U}}, \label{X}\\
&& \{p, {\bar x}\}=\{\bar{p}, x\}=1, \quad \left\{ p, \bp \right\} = \{p, x\}= \{\bar{p}, {\bar x}\}= \{x, \bar{x}\}=0 \label{PB2}.
\eea
Hence, we have shown that the deformed free particle introduced in \cite{KLS} is canonically equivalent to the ordinary free particle.
Respectively, the actions of both systems  admit $SU(1,2)$ invariance, which is reduced in the $\gamma=0$ limit to
the $\ell=1/2$-extended conformal Galilei group.

It is  instructive to write explicit realization of the generators of the  $su(1,2)$ algebra in terms of the canonical variables
$x, {\bar x}, p, \bp$:
\be\label{realis}
\begin{aligned}
&L_{-1}=\mathcal{H}_0 = p\, \bar{p},\quad  L_0  =-\frac{1}{2} \left(p\,\bar{x}+\bar{p}\, x\right),\quad
L_1 =x \, \bar{x}, \\
& U= \im \left(x\,\bar{p} - \bar{x}\,p\right), \quad
G_{-1/2}=-p \sqrt{1+ \gamma U}\,, \; \bG_{-1/2}=- \bp \sqrt{1+ \gamma U}, \\
& G_{1/2}=x \sqrt{1+ \gamma U} ,\; \bG_{1/2}=\bx \sqrt{1+ \gamma U} .
\end{aligned}
\ee
Time-dependent extensions of these generators,  defining the isometries of the Lagrangian, are given by the expressions
\be
\begin{aligned}
&L^t_{-1}=L_{-1},\quad L^t_0=L_0+t\,L_{-1},\quad L^t_1=L_1+2 t \,L_0+t^2\, L_{-1},\quad U^t = U,  \\
 &G^t_{-1/2}=G_{-1/2},\quad \bG^t_{-1/2}=\bG_{-1/2},\quad G^t_{1/2}=G_{1/2}+t\,G_{-1/2},\quad \bG^t_{1/2}=\bG_{1/2}+t\,\bG_{-1/2}.
\end{aligned}
\ee
The respective Hamiltonian vector fields  restricted to the second-order Lagrangian  (parameterized by  $x,\bar{x}$  ) define the following symmetry  transformations
\be
\label{vec}
\begin{aligned}
&{\bf L}_{-1}=-\dot{x}\frac{\partial}{\partial x}+\quad c.\;c. \\
&{\bf L}_{0}=\left(-\frac12x+  2t\dot{x}\right)\frac{\partial}{\partial x}+\quad c.\;c.  \\
&{\bf L}_{1}=\left(-t x+  t^2\dot{x}\right)\frac{\partial}{\partial x}+\quad c.\;c.   \\
&{\bf {G}}_{-1/2}=-\frac{1+\imath\gamma\bar{x} \dot{x}  -\frac{3\imath}{2}\gamma \bar{x} \dot{\bar{x}}   }{\sqrt{  1+\imath\gamma   \bar{x} \dot{x}
-\imath\gamma {x} \dot{\bar{x}}   }}\frac{\partial}{\partial x}
-   \frac{    \imath\gamma\bar{x}\dot{\bar{x}}}{2\sqrt{  1+\imath\gamma   \bar{x} \dot{x}
 -\imath\gamma {x} \dot{\bar{x}} }  }  \frac{\partial}{\partial\bar{x}}  \\
& {\bf G}_{1/2}=-\frac{2t+\imath\gamma\bar{x} {x}  +2\imath\gamma t \bar{x} \dot{{x}}  -   3\imath\gamma t{x} \dot{\bar{x}}         }{2\sqrt{  1+\imath\gamma   \bar{x} \dot{x}
-\imath\gamma x \dot{\bar{x}}    }}\frac{\partial}{\partial x}
+   \frac{    \imath\gamma\bar{x}({\bar{x}}  -t\dot{\bar{x}}) }{2\sqrt{  1+\imath\gamma   \bar{x} \dot{x}
 -\imath\gamma {x} \dot{\bar{x}} }  }  \frac{\partial}{\partial\bar{x}}
 \end{aligned}
\ee

It is worth noting that the explicit realization of the $su(1,2)$ algebra \p{realis} makes evident the statement that
the $su(1,2)$ algebra as well as its $\gamma=0$ reduction (i.e. the $\ell=1/2$ conformal Galilei algebra) can be constructed in terms of
two oscillators $(x,\bp)$ and $(\bx,p)$. Thus, both these algebras lie in the enveloping algebra of two oscillators.

\subsection{Oscillator}
Now, let us  consider the general case of deformed oscillator given by the Hamiltonian \eqref{H} which is simply $L_{-1}+\omega^2 L_{1}$.
In addition to constant of motion $U$ given in \eqref{simfree} defining  $U(1)$ symmetry,   it possesses hidden symmetries
given by the generalization of Fradkin tensor
\be
\begin{aligned}
&A = {G^2_{-1/2}}+ \frac{1}{\omega^2}\left(\{ \mathcal{H}, G_{-1/2}\}\right)^2= {G^2_{-1/2}}+ \omega^2\, \left( u+\gamma u U-\frac{\im}{2} \gamma  \bu u {G_{-1/2}} \right)^2\\
&{\bar A} = {\overline G}^2_{-1/2}+ \frac{1}{\omega^2}\left(\{ \mathcal{H}, {\overline G}_{-1/2}\}\right)^2={\overline G}^2_{-1/2} + \omega^2\,
\left( \bu+\gamma \bu  U+\frac{\im}{2} \gamma u\bu {\overline G}_{-1/2} \right)^2.
\end{aligned}
\ee
These constants of motion form the deformation of $su(2)$ algebra
\be
\{A, {\bar A}\}=-4\im\omega^2(U+3\gamma U^2 +2\gamma^2 U^3) +
  4 \im \gamma (1+\gamma U)\mathcal{H}^2, \quad \{U, A_\pm\}=\mp 2 \im A_\pm.
\ee
The Hamiltonian is the Casimir of this algebra. It expresses via  above constants of motion as follows
\be
\mathcal{H}^2={\frac{A\, {\bar  A}}{\left(1+\gamma U\right)^2}+\omega^2 U^2}.
\label{fr0}\ee
One may  may directly check
that the Hamiltonian \p{H}, being rewritten in terms of  canonical variables $x, \bx, p, \bx$ \p{pp},\p{X}, acquires the form
\be\label{Ham3}
\mathcal{H}= p\, \bp + \omega^2 x\,\bx
\ee
as it should be.
To get the canonical formulation of symmetry algebra, we redefine the Fradkin tensors as
\be
\mathcal{A}=\frac{A}{1+\gamma U} = 
{p^2}+\omega^2 x^2,\quad
\overline{\mathcal{A}}=\frac{{\bar A}}{1+\gamma U} =
{\bar{p}^2}+ \omega^2\; \bar{x}^2\; :\qquad\{ {\cal A},{\overline{\cal A}}\}=-4 \im \omega^2 U,\quad \{U, \mathcal{A}\}=- \,2\, \im\, \mathcal{A}.
\ee
where $p, {\bar p} ,x, \bar{x} $ are given  by \p{pp}, \p{X}.
%
%
In terms of these tensors the Hamiltonian of the oscillator reads
\be
\mathcal{H}{}^2= {\cal A}\,{\overline{\cal A}} +\omega^2 U^2.
\ee
Hence, the deformed oscillator is canonically (classically) equivalent to non-deformed one.  Since its action admits the invariance under $su(1,2)$ algebra, we conclude that non-deformed oscillator action possesses  the same invariance  as well.
\section{Conclusion}
In this paper, we provided the Hamiltonian description of the deformed two dimensional oscillator \cite{KLS}, i.e. oscillator possessing
dynamical $SU(1,2)$ symmetry. One of the interesting features of this system is the fact that its first-order Lagrangian
is nothing but one of the Cartan forms defined on the coset $SU(1,2)/H$ with a quite unusual choice of the stability
subgroup $H$, which includes the dilatation and conformal boosts together with $U(1)$ rotation. On the other hand, this one-form
is a source of the symplectic form and the Hamiltonian, both written in terms of the initial variables. In this basis, the
Hamiltonian of the deformed oscillator is simple, while the Poisson brackets are more involved. Analysing the structure of the
Hamiltonian we succeeded in finding the canonical variables in which the Hamiltonian of the deformed oscillator coincides with
the Hamiltonian of the ordinary two dimensional oscillator. Thus, we proved that these two systems, deformed and ordinary
two dimensional oscillators, are canonically equivalent.

Proving the canonical equivalence of these systems, we have explicitly constructed the currents spanning $su(1,2)$ algebra in terms
of the ordinary oscillator. The main feature of this realization is a non-polynomial structure of $su(1,2)$ currents. Probably just this
property was the obstacle preventing from immediate visualization of the $su(1,2)$ algebra within the enveloping algebra of the two-dimensional
oscillator.

The established equivalence of the deformed and ordinary oscillators within the Hamiltonian approach does not mean their equivalence
as the Lagrangian systems. Indeed, the transformations between these two systems are forbidden at the Lagrangian level. Thus,
the Hamiltonian formulation is  more suitable for analysis of this type of systems, as compared to the Lagrangian one.

Concerning the further developments, one has to note that the  $su(1,2)$ algebra is not a  unique one which admits reduction to
the conformal Galilei algebra and thus can be viewed as its deformation. The immediate example of possible algebras having the proper
structure is provided by the wedge subalgebras in the $U(n)$ quasi-superconformal algebras \cite{LJR}. A preliminary analysis shows
that the extension of the approaches of \cite{KLS} and the Hamiltonian formalism of the present paper to these algebras will result
in the $SU(n+2)$ invariant $d=n+1$ dimensional oscillators. Another interesting algebra is $so(2,3)$ which may be viewed as
a deformation of the three-dimensional $\ell=1$ conformal Galilei algebra \cite{KNS}.
\section*{Acknowledgements}
We are grateful to Anton Galajinsky, Alexander Sorin and Mikhail Vasiliev for valuable correspondence.
The work of S.K. was partially supported by RSCF, grant 14-11-00598 and by
RFBR, grant 15-52-05022 Arm-a.
The work of A.N. was  partially supported by the Armenian
State Committee of Science, Grants No. 15RF-039 and No. 15T-1C367, and  it was done within ICTP programs NET68 and OEA-AC-100.


\begin{thebibliography}{99}
\bibitem{confNH1} P.~Havas, J.~Plebanski, \\
{\it Conformal extensions of the Galilei group and their relation to the Schr\"{o}dinger group}, \\
J.~Math.~Phys. {\bf 19} (1978) 482.
\bibitem{confNH2}J.~Negro, M.A.~del~Olmo, A.~Rodriguez-Marco, \\
{\it Nonrelativistic conformal groups}, \\
J.~Math.~Phys. {\bf 38} (1997) 3786.
\bibitem{confNH3} A.~Galajinsky, I.~Masterov, \\
{\it Remarks on l-conformal extension of the Newton--Hooke algebra}, \\
Phys.~Lett. {\bf B702} (2011) 265, {\tt [arXiv:1104.5115[hep-th]]}.
\bibitem{confPU}K.~Andrzejewski, A.~Galajinsky, J.~Gonera, I.~Masterov,\\
{\it Conformal Newton-Hooke symmetry of Pais--Uhlenbeck oscillator}, \\
Nucl.~Phys. {\bf B885} (2014) 150, {\tt [arXiv:1402.1297[hep-th]]}.
\bibitem{GM1}A.~Galajinsky, I.~Masterov, \\
{ \it Dynamical realizations of l-conformal Newton--Hooke group},\\
Phys.~Lett. {\bf B723} (2013) 190, {\tt [arXiv:1303.3419[hep-th]]}.
\bibitem{Ivanov}S.~Fedoruk, E.~Ivanov, J.~Lukierski, \\
{\it Galilean conformal mechanics ,from nonlinear realizations},\\
Phys.~Rev. {\bf D83} (2011) 085013 {\tt [arXiv:1101.1658[hep-th]]}.
\bibitem{MT} D.~Martelli, Y.~Tachikawa, \\
{\it Comments on Galilean conformal field theories and their geometric realization},\\
JHEP {\bf 1005} (2010) 091, {\tt [arXiv:0903.5184[hep-th]]}.
\bibitem{PU11} K.~Andrzejewski, \\
{\it Hamiltonian formalisms and symmetries of the Pais--Uhlenbeck oscillator},\\
Nucl.Phys. {\bf B889} (2014) 333,  {\tt arXiv:1410.0479[hep-th]}.
\bibitem{coset1} S.R.~Coleman, J.~Wess, B.~Zumino,\\
{\it Structure of phenomenological lagrangians. 1},\\
Phys.\ Rev.\ {\bf 177} (1969) 2239;\\
C.G.~Callan, S.R.~Coleman, J.~Wess, B.~Zumino,\\
{\it Structure of phenomenological lagrangians. 2},\\
Phys.\ Rev.\ {\bf 177} (1969) 2247.
\bibitem{coset2} D.V.~Volkov,\\
{\it Phenomenological lagrangians},\\
Sov.\ J.\ Part.\ Nucl.\ {\bf 4} (1973) 3; \\
V.I.~Ogievetsky,\\
{\it Nonlinear realizations of internal and space-time symmetries},\\
in: Proceedings of the Xth Winter School of Theoretical Physics in Karpacz, Vol.1, p.117, 1974.
\bibitem{ih} E.A.~Ivanov, V.I.~Ogievetsky,\\
{\it The inverse Higgs phenomenon in nonlinear realizations},\\
Teor.\ Mat.\ Fiz.\ {\bf 25} (1975) 164.
\bibitem{KLS} S.~Krivonos, O.~Lechtenfeld, A.~Sorin, \\
{\it Minimal realization of $\ell$-conformal Galilei algebra, Pais-Uhlenbeck oscillators and their deformation},\\
{\tt arXiv:1607.03756[hep-th]}.
\bibitem{LJR} L.J.~Romans,\\
{\it Quasi-superconformal algebras in two dimensions and Hamiltonian reduction}, \\
Nucl. Phys. {\bf B357} (1991) 549.
\bibitem{KNS} S.Krivonos, A.~Nersessian, A.~Sutulin,
{\it In preparation}.
\end{thebibliography}
\end{document}